\pdfoutput=1

\documentclass[11pt]{article}

\usepackage[preprint]{acl}

\usepackage{times}
\usepackage{latexsym}

\usepackage[T1]{fontenc}

\usepackage[utf8]{inputenc}

\usepackage{microtype}

\usepackage{inconsolata}

\usepackage{graphicx}
\graphicspath{{../}{./}}
\usepackage{algorithm}
\usepackage{algpseudocode}

\usepackage{amssymb}
\usepackage{amsmath}
\usepackage{booktabs}
\usepackage{multirow}
\usepackage{caption}
\usepackage[dvipsnames]{xcolor}
\usepackage[table]{xcolor} 
\usepackage{booktabs}      
\usepackage{graphicx}
\usepackage{tablefootnote}

\renewcommand{\algorithmicrequire}{\textbf{Input:}}
\renewcommand{\algorithmicensure}{\textbf{Output:}}
\renewcommand{\algorithmicloop}{\textbf{repeat}}

\usepackage{enumitem}
\usepackage{booktabs}
\usepackage{multirow}
\usepackage[dvipsnames]{xcolor}
\usepackage{pifont} 
\usepackage{graphicx}
\usepackage{threeparttable}
\usepackage{pgfplots}
\usepackage{booktabs}   
\usepackage{subcaption}
\usepackage{tabularx}  
\usepackage{makecell}
\usepackage{array}      
\usepackage{pgfplots}
\pgfplotsset{compat=1.18} 
\usetikzlibrary{patterns}  
\usepackage{tcolorbox}
\usepackage{hhline}
\usepackage{fontawesome5} 
\usepackage{amsmath}
\usepackage{dashrule} 
\usepackage{placeins}
\usepackage{algorithm}
\usepackage{mathtools} 
\usepackage[noEnd=true]{algpseudocodex}
\usepackage{footmisc}
\usepackage{tablefootnote}
\tikzset{algpxIndentLine/.style={draw=gray,very thin}}

\algrenewcommand\algorithmicrequire{\textbf{Input:}}
\algrenewcommand\algorithmicensure{\textbf{Output:}}
\algrenewcommand\algorithmicloop{\textbf{repeat}}
\usepackage{amsmath}
\tcbuselibrary{skins, breakable}

\definecolor{LightGray}{gray}{0.95} 
\definecolor{LightBlue}{rgb}{0.92, 0.95, 1.0} 
\newcommand{{\method}}{R$^2$A}
\newtcolorbox{gadgetbox}{
  breakable,
  colback=red!3,
  colframe=black!80,
  boxrule=0.8pt,
  arc=0.5mm,
  left=6pt, right=6pt, top=6pt, bottom=6pt,
  fontupper=\small,
  before skip=2pt,
  after skip=4pt
}

\title{Route to Rome Attack: Directing LLM Routers to Expensive Models via Adversarial Suffix Optimization}

\author{
  \textbf{Haochun Tang}\textsuperscript{1,2}\thanks{Equal contribution, co-first author.},
  \textbf{Yuliang Yan}\textsuperscript{2}\footnotemark[1],
  \textbf{Jiahua Lu}\textsuperscript{1,2},
  \textbf{Huaxiao Liu}\textsuperscript{1}\thanks{Corresponding author.},
  \textbf{Enyan Dai}\textsuperscript{2}\footnotemark[2] \\
\\
 \textsuperscript{1}Key Laboratory of Symbolic Computation and Knowledge Engineering, MoE, Jilin University\\
 \textsuperscript{2}The Hong Kong University of Science and Technology (Guangzhou)\\
\\
 \small{
    \href{tanghc24@mails.jlu.edu.cn}{tanghc24@mails.jlu.edu.cn},
    \href{enyandai@hkust-gz.edu.cn}{enyandai@hkust-gz.edu.cn}
 }
}

\begin{document}
\maketitle
\begin{abstract}
Cost-aware routing dynamically dispatches user queries to models of varying capability to balance performance and inference cost. However, the routing strategy introduces a new security concern that adversaries may manipulate the router to consistently select expensive high-capability models. Existing routing attacks depend on either white-box access or heuristic prompts, rendering them ineffective in real-world black-box scenarios. In this work, we propose R$^2$A, which aims to mislead black-box LLM routers to expensive models via adversarial suffix optimization. Specifically, R$^2$A deploys a hybrid ensemble surrogate router to mimic the black-box router. A suffix optimization algorithm is further adapted for the ensemble-based surrogate. Extensive experiments on multiple open-source and commercial routing systems demonstrate that {R$^2$A} significantly increases the routing rate to expensive models on queries of different distributions. Code and examples: \url{https://github.com/thcxiker/R2A-Attack}.

\end{abstract}

\section{Introduction}
The development of Large Language Models (LLMs) has achieved remarkable success. 
These improvements are fundamentally driven by scaling laws~\citep{scaling}, which indicate that performance improves predictably with increased model size. 
For instance, Qwen-3-Max scales to over 1 trillion parameters, approximately $14\times$ larger than the previous flagship Qwen-2.5-72B. 
However, serving every user query with such state-of-the-art models is computationally and economically unsustainable for commercial adoption. 

\begin{figure}[t]
    \centering
    \includegraphics[width=0.95\linewidth]{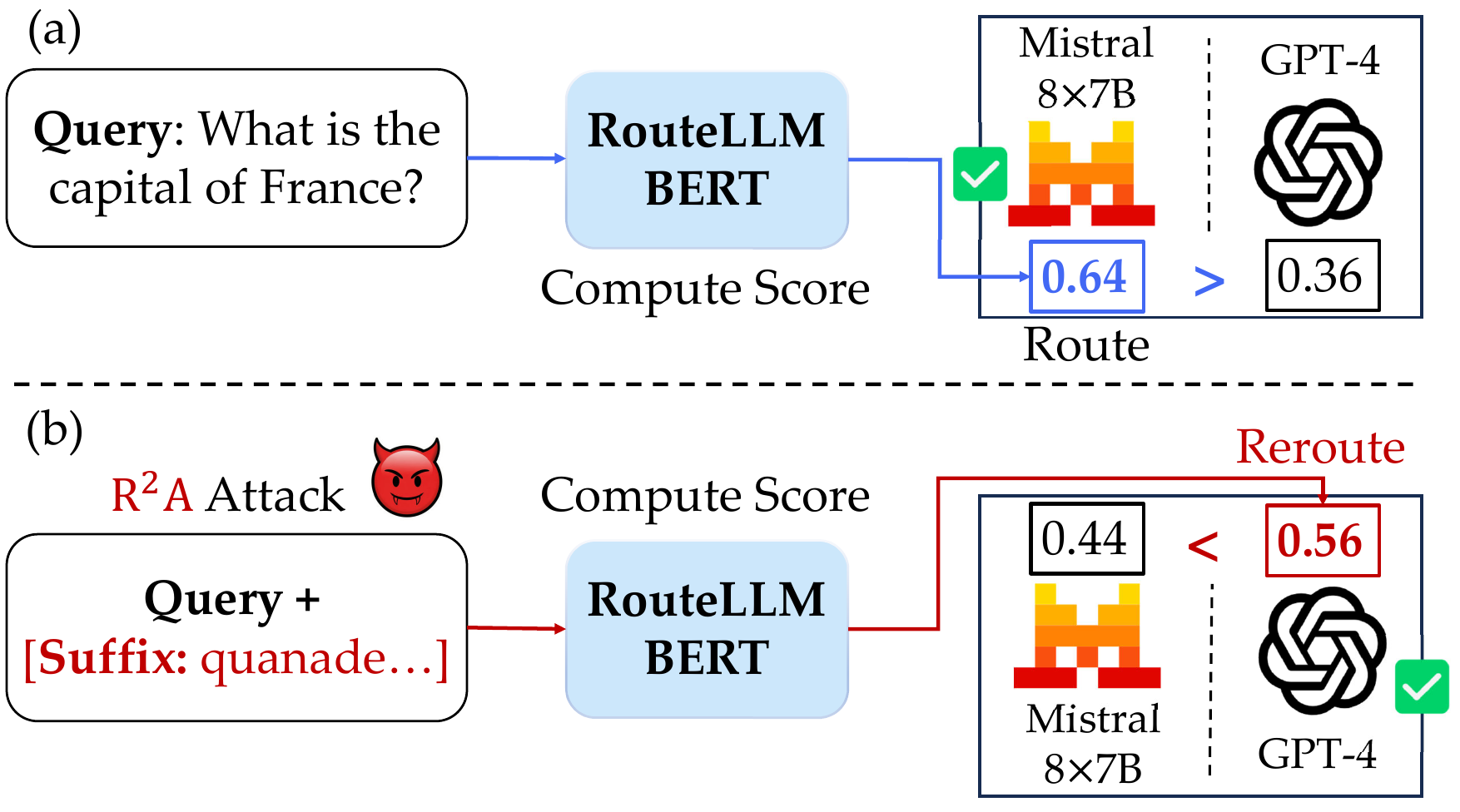} 
    \caption{ (a) An example of cost-aware LLM routing. (b) The corresponding routing attack by our {\method}.}
    \label{fig:router_concept}
    \vspace{-1em}
\end{figure}

To balance performance and cost, {cost-aware LLM routing} has been proposed to route each query to the least-cost model that meets a target quality~\citep{zooter,aggarwal2025automixautomaticallymixinglanguage,routellm}. 
This strategy is grounded in the insight that only a small fraction of requests necessitate expensive strong models, whereas simple queries can be effectively handled by cheaper weak models.
As illustrated in Fig.~\ref{fig:router_concept}(a), upon receiving the simple factual query ``What is the capital of France?'', the router identifies it as low-complexity and selects the weak Mistral 8x7B model to generate the response.
Such routing has also been adopted in commercial systems such as OpenRouter~\footnote{\url{https://openrouter.ai/openrouter/auto/}\label{fn:openrouter}} and GPT-5-Auto~\footnote{\url{https://openai.com/index/gpt-5-system-card/}\label{fn:gpt5}}. 

Despite its effectiveness, cost-aware routing raises a natural concern of routing attack: \textit{Can an adversary use a universal trigger (e.g., a fixed suffix) to consistently manipulate the router toward expensive models?} 
Some initial investigations have been conducted to answer this question~\citep{reroutingllmrouters,lin2025life}. However, \citet{reroutingllmrouters} relies on accessible gradients or known architectures of target routers. 
This is impractical in commercial settings where black-box routers only allow observing the final routing decision. 
As for LifeCycle~\citep{lin2025life}, it extracts templates like ``Below is an instruction..., [query]'' from high-win-rate queries to guide the router to select expensive LLMs. 
While applicable in black-box settings, such heuristic prompts are not rigorously optimized and are therefore often insufficient to consistently manipulate various target routers. 

Therefore, in this work, we introduce \underline{R}oute to \underline{R}ome \underline{A}ttack ({\method}), which optimizes suffixes with only black-box access to the target router. 
As the example in Fig.~\ref{fig:router_concept}(b), after appending the learned suffix to the query, the target router would reroute the simple query to expensive strong models. 

Inspired by previous black-box attack methods~\citep{surrogate2,surrogate1}, {\method} trains a surrogate router to mimic the target router, enabling optimization of the adversarial suffix. 
However, two key challenges remain to be addressed: (i) how to build a faithful surrogate router in a black-box setting with a strict query budget? Existing routers utilize diverse mechanisms, varying from semantic embeddings to LLM-based approaches. 
Without knowledge of the target router's architecture, it is challenging to learn a surrogate router that faithfully mimics the target router's behavior. 
Moreover, the query budget constraint further complicates surrogate construction. 
(ii) Even with a surrogate router, it remains challenging to optimize a discrete adversarial suffix for effective router attacks across diverse queries.

To address the above two challenges, {\method} introduces a hybrid ensemble surrogate router $\mathcal{R}_s$ that combines diverse routing mechanisms including multiple existing open-source methods and lightweight trainable routers. 
By ensembling multiple router architectures, the surrogate better aligns with unknown target routers $\mathcal{R}_t$ under limited query budgets. Additionally, we propose a suffix optimization algorithm designed specifically to aggregate gradients effectively for the ensemble surrogate. Experiments on 6 datasets across 7 open-source and 2 commercial black-box routers (GPT-5-Auto and OpenRouter), demonstrate that {\method} effectively optimizes adversarial suffixes to mislead routers to expensive models. Our main contributions can be summarized as:
\begin{itemize}[leftmargin=*, itemsep=0pt,topsep=0pt]
    \item We study a novel problem of directing black-box LLM routers to expensive models via adversarial suffix optimization;
    \item Our proposed {\method} introduces a novel hybrid ensemble surrogate router to mimic the router within limited black-box queries, along with a tailored adversarial suffix optimization algorithm.
    \item Extensive experiments validate that {\method} effectively generalizes to diverse routers, including commercial GPT-5-{Auto} and OpenRouter.
\end{itemize}

\section{Problem Definition}
In this section, we formalize the problem of attacking LLM routers in a realistic black-box setting.

\subsection{Preliminaries of LLM Router}
Given a query $q$, a LLM router $\mathcal{R}: q\rightarrow \mathbb{R}^N$ selects a model from a pool $\mathcal{M}=\{M_1,\dots,M_N\}$. 
For cost-aware routing, the router aims to minimize inference cost while meeting a target quality constraint by solving~\citep{routellm}:
\begin{equation}
\mathcal{R}(q)=\arg\min_{M_i\in\mathcal{M}}
\Big({\ell}(q,M_i)+\lambda\cdot {C}(q,M_i)\Big),
\end{equation}
where ${\ell}(q,M_i)$ denotes the predicted loss of model $M_i$ on $q$,
${C}(q,M_i)$ is the cost score, and $\lambda\ge 0$ controls the contribution of cost score in routing.

\subsection{Threat Model of Router Attack}

\textbf{Attacker's Goal}. The goal is to mislead the router into selecting expensive models to answer the given query.  Specifically, following~\citet{reroutingllmrouters}, we partition model candidate pool into expensive strong models $\mathcal{M}_{\text{strong}}$ and cheap weak models $\mathcal{M}_{\text{weak}}$ using public leaderboards\footnote{\url{https://lmarena.ai/leaderboard}}. $\mathcal{M}_{\text{strong}}$ incurs substantially higher inference cost than $\mathcal{M}_{\text{weak}}$, which is described in Appendix~\ref{app:Model Partition}.
Formally, given a query $q$ such that 
$\mathcal{R}_t(q)\in\mathcal{M}_{\text{weak}}$, 
an router attack operation $\mathcal{A}$ succeeds if target router $\mathcal{R}_t(\mathcal{A}(q))\in\mathcal{M}_{\text{strong}}$.

\noindent \textbf{Attacker's Capability}. The attacker can modify the original query by appending an adversarial {suffix}. To preserve answer quality and keep the modification minimal, the attacker is restricted to appending a suffix $s$ of at most $\Delta$ tokens to the end of the query $q$.

\noindent \textbf{Attacker's Knowledge}. 
We assume a realistic black-box setting where the attacker can only observe the target router's decision for an input query. 
As shown in Table~\ref{tab:router_transparency}, this assumption aligns with current commercial practices where routing services typically expose their candidate model pools and selected model decisions to ensure billing transparency.  
All other information, such as the target router's internal logits, parameters, or gradients, is inaccessible. Because each query to the target router generally incurs a financial cost, the attacker is restricted to at most $Q$ queries to $\mathcal{R}_t$.
For GPT-5-Auto, where routing decisions are not observable, we apply the suffix learned on an OpenRouter.

\subsection{Router Attack Formulation}
\textit{Our objective is to find a universal adversarial suffix $s^*$ that can alter the decision of the target router $\mathcal{R}_t$ to expensive strong models} . Given the above threat model, the router attack can be formulated as an optimization problem where the attacker seeks a suffix $s^*$ that maximizes the expected probability of routing a query to a strong model by:
\begin{equation}\label{eq:target}
    \begin{aligned}
    s^* =  & \underset{s}{\arg \max}  \; \mathbb{E}_{q \sim \mathcal{Q}} \left[\mathbb{I}(\mathcal{R}_t(q\oplus s) \in \mathcal{M}_{\text{strong}}) \right] \\
    &  ~~ \text{s.t.} \quad s \in \mathcal{S}, \quad |s| \leq \Delta,
    \end{aligned}
\end{equation}
where $\mathcal{Q}$ denotes the distribution of input queries, $\oplus$ represents the concatenation operation, $\mathbb{I}(\cdot)$ is the indicator function, and $\Delta$ specifies the maximum token length budget for the adversarial suffix.



\label{sec:method}


\section{Method}
\begin{figure*}[!ht]
    \centering
    \includegraphics[width=0.85\linewidth]{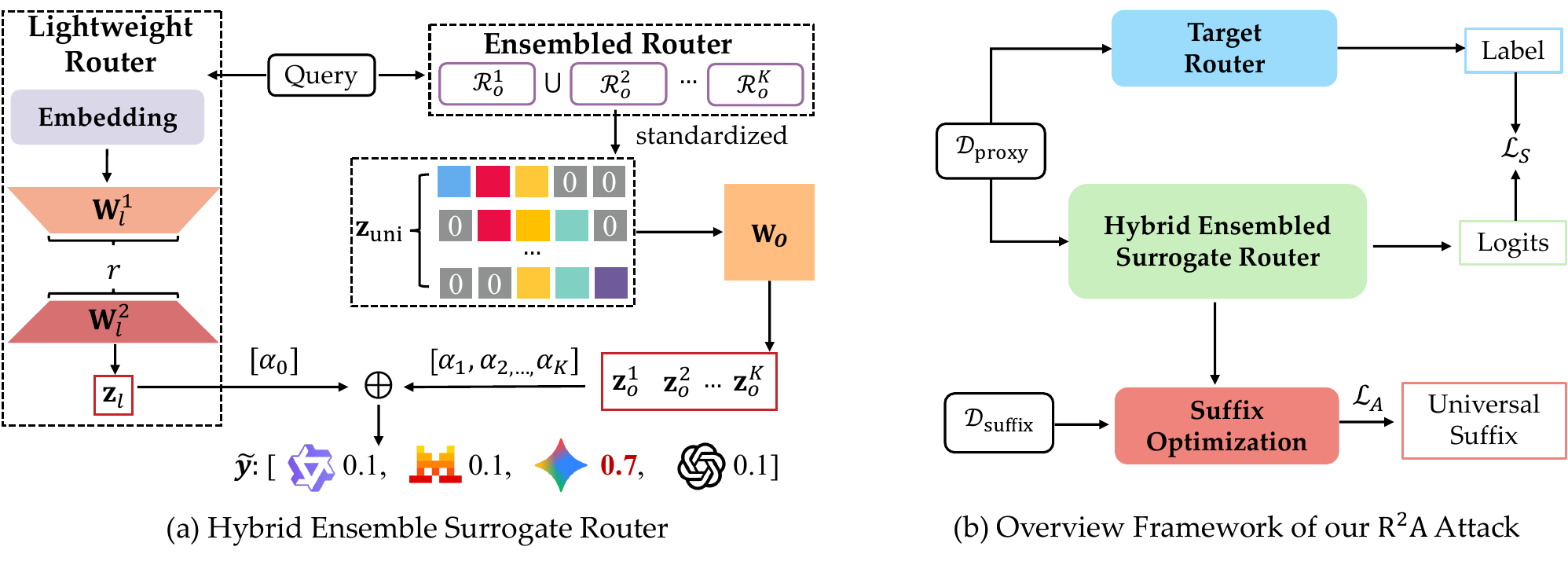}
    \vspace{-0.5em}
    \caption{Framework of our \method{}. (a) We design a hybrid ensemble surrogate router, including a lightweight router and an ensembled router. (b) Our {\method} pipeline consists of surrogate model training, followed by suffix optimization.}
    \vspace{-1em}
    \label{fig:overview}
\end{figure*}

Under the black-box setting, we have no access to its parameters or gradients of the target router $\mathcal{R}_t$. Therefore, Eq.(\ref{eq:target}) cannot be directly optimized via gradient descent. 
Therefore, {\method} first trains a surrogate router to mimic the target router’s behavior, and then uses the surrogate router to optimize a universal adversarial suffix.
As shown in Fig.~\ref{fig:overview}, {\method} introduces a hybrid ensemble surrogate router that combines diverse existing open-source routers with lightweight trainable routers. 
By covering diverse routing mechanisms, the surrogate can better align with the target router $\mathcal{R}_t$ of unknown design within the query budget. 
In addition, a suffix optimization algorithm is further adopted for the hybrid ensemble surrogate router.
Next, we introduce each component of {\method} in detail. 




\subsection{Hybrid Ensemble Surrogate Router}
\label{sec:surrogate_construction}
Since the target router design is unknown, relying on a single architecture may cause an architectural mismatch and thus yield a poor surrogate. 
Hence, we build a {hybrid ensemble surrogate router} that combines diverse pre-trained open-source routers and trainable lightweight routers. 
During the surrogate training, we jointly learn the ensemble weights of all routers and the parameters of the lightweight router. 
This design offers two key advantages:
(i) By incorporating open-source routers, {\method} can quickly identify an existing router (or a linear combination) that matches the target behavior, reducing the required queries;
(ii) By optimizing a trainable lightweight router, {\method} can handle target routers significantly different from all pre-trained open-source routers. 
Next, we introduce details of the hybrid ensemble surrogate router. 

\noindent \textbf{Design of Trainable Lightweight Router}. This trainable lightweight router $\mathcal{R}_l$ aims to predict the target router's decision from the query's embedding ${E}(q) \in \mathbb{R}^{d}$.
Specifically, \texttt{all-MiniLM-L6-v2}~\cite{MINILM} is deployed as the encoder, where $d=384$. 
However, directly learning a linear mapping $\mathbb{R}^{d}\rightarrow \mathbb{R}^{|\mathcal{M}_t|}$ involves optimizing a parameter matrix of size $d\times |\mathcal{M}_t|$, where $|\mathcal{M}_t|$ denotes the number of candidate models in the target router. 
Training this large matrix demands extensive queries, exceeding the strict query budget. 
Inspired by LoRA~\citep{lora}, we impose a low-rank constraint by decomposing the transformation into two smaller matrices 
$\mathbf{W}_l^1 \in\mathbb{R}^{d\times r}$ and 
$\mathbf{W}_l^2 \in\mathbb{R}^{r\times |\mathcal{M}_{\text{t}}|}$, 
with rank $r \ll d$. 
The logits from the lightweight router $\mathbf{z}_l\in \mathbb{R}^{|\mathcal{M}_{\text{t}}|}$ are then computed as:
\begin{equation} \label{eq:lightlogits}
\mathbf{z}_l = {E}(q)\mathbf{W}_l^1\mathbf{W}_l^2,
\end{equation}
With this low-rank decomposition, fewer queries are required for training the router $\mathcal{R}_l$.

\noindent \textbf{Combining with Open-Source Routers}. {\method} combines multiple open-source routers with diverse routing mechanisms, i.e., $\{\mathcal{R}_o^{(1)},\dots,\mathcal{R}_o^{(K)}\}$. 
This would reduce the mechanism mismatch between the surrogate router and target router. 
However, the model pools of open-source routers $\{\mathcal{M}_o^{(1)},\dots,\mathcal{M}_o^{(K)}\}$ are inconsistent with each other and the target router's model pool $\mathcal{M}_t$. 
Hence, we map their logits to the union of all open-source model pools, i.e., $\mathcal{M}_{\text{uni}} = \bigcup_{k=1}^{K} \mathcal{M}_o^{(k)}$. {Zero-padding} is applied to handle missing candidates. 
Formally, for an open-source router $\mathcal{R}_o^{(k)}$, we extend its logit vector as ${\mathbf{z}}^{(k)}_{\text{uni}} = [\tilde{z}^{(k)}_1, \dots, \tilde{z}^{(k)}_{|\mathcal{M}_{\text{uni}}|}]$, where each element is defined as:
\begin{equation}
\tilde{z}^{(k)}_i =
\begin{cases}
z^{(k)}_{M_i}, & \text{if } M_i \in \mathcal{M}_o^{(k)}, \\
0, & \text{otherwise},
\end{cases}
\end{equation}
where $\tilde{z}^{(k)}_i$ is the standardized logit for model $M_i \in \mathcal{M}_{\text{uni}}$, and $z^{(k)}_{M_i}$ is the original logit value assigned to $M_i$ by the open-source router $\mathcal{R}_k^o$.

As the union model pool of open-source routers $\mathcal{M}_{\text{uni}}$ typically differs from the target pool $\mathcal{M}_t$, we apply a linear mapping to align their logits:
\begin{equation} \label{eq:openlogits}
    \mathbf{z}^{(k)}_{\text{o}} = \mathbf{W}_o \cdot \mathbf{z}^{(k)}_{\text{uni}},
\end{equation}
where $\mathbf{W}_o \in \mathbb{R}^{|\mathcal{M}_{\text{uni}}|\times |\mathcal{M}_{t}|}$ is the projection matrix, and $\mathbf{z}^{(k)}_{\text{o}}$ denotes the logits of router $\mathcal{R}_o^{(k)}$projected onto the target router's model pool space.

With Eq.(\ref{eq:openlogits}) and Eq.(\ref{eq:lightlogits}), we can get the prediction logits of open-source routers and lightweight trainable routers, respectively. 
Then, the ensembles' routing results on the target model pool $\mathcal{M}_t$ can be computed by a weighted summation:
\begin{equation}
    \hat{y} = \text{softmax}(\alpha_{0} \mathbf{z}_l  + \sum_{i=1}^{K} \alpha_i \mathbf{z}_o^{(k)}),
\end{equation}
where $\alpha_i$ are learnable ensemble weights satisfying $\alpha_i \geq 0$ and $\sum_{i=0}^{K}\alpha_i=1$.

\noindent \textbf{Surrogate Router Training.} To train the surrogate router, we query the black-box target $\mathcal{R}_{\text{target}}$ to generate training labels. According to the threat model, we are limited to querying the target router $Q$ times. The surrogate training objective optimizes parameters $\theta=\{\mathbf{W}_l^1, \mathbf{W}_l^2, \mathbf{W}_o, \{\alpha_i\}_{i=0}^{K}\}$ by minimizing:
\begin{equation}
\min_{\theta} \mathcal{L}_{S} = \frac{1}{Q} \sum_{i=1}^{Q} l(\hat{y}(q_i), \mathcal{R}_t(q_i)),
\end{equation}
where $l(\cdot)$ is the cross-entropy loss, and $\hat{y}(q_i)$ and $\mathcal{R}_t(q_i)$ represent the predictions of surrogate and target router for query $q_i \in \mathcal{D}_{\text{proxy}}$, respectively.

\subsection{Adversarial Suffix Optimization with Hybrid Ensemble Surrogate Router}
\label{sec:adversarial_optimization}
With the hybrid ensemble surrogate router, adversarial suffix optimization can be reformulated as:
\begin{equation} \label{eq:attack_obj}
    \min_{s} \mathcal{L}_{A} = - \mathbb{E}_{q\sim \mathcal{Q}} \sum_{M \in \mathcal{M}_{\text{strong}}} p(\hat{y}=M|q \oplus s), 
\end{equation}
where $p(\hat{y}=M|q \oplus s)$ denotes the surrogate router's predicted probability that the query $q$ appended with adversarial suffix $s$ is routed to model $M$. 
One may deploy Greedy Coordinate Gradient (GCG)~\cite{GCG}, which greedily replaces tokens using token gradients from the encoder.
However, our ensemble surrogate involves multiple encoders. 
Hence, gradient aggregation across routers in the ensemble surrogate is required. 

\noindent \textbf{Aggregation of Suffix Token Gradients}. We first analyze the token gradient via the chain rule. 
Let $\mathbf{z}_{\text{total}} = \sum_{k=0}^K \alpha_k \mathbf{z}^{(k)}$ denote the ensemble logits. 
Consequently, for the $k$-th router, the gradient w.r.t the token $s_i$ is calculated as:
\begin{equation}
    {g}^{(k)}_{i} = \frac{\partial \mathcal{L}_A}{\partial \mathbf{z}_{\text{total}}} \cdot \underbrace{\frac{\partial \mathbf{z}_{\text{total}}}{\partial \mathbf{z}^{(k)}}}_{\alpha_k} \cdot \frac{\partial \mathbf{z}^{(k)}}{\partial s_i} 
    = \alpha_k \cdot \frac{\partial \mathcal{L}_A}{\partial \mathbf{z}_{\text{total}}} \frac{\partial \mathbf{z}^{(k)}}{\partial s_i}.
\label{eq:gradient_chain}
\end{equation}
For the term $\delta_i^{(k)}=\frac{\partial \mathbf{z}^{(k)}}{\partial s_i}$ in Eq.(\ref{eq:gradient_chain}), while it effectively captures the token sensitivity within a single router, its magnitude can vary drastically across different architectures. 
Therefore, direct summation of ${g}_i^{(k)}$ would lead to a specific member router dominating the optimization. 
To mitigate this bias, we normalize the term $\delta_i^{(k)}$ to the range $[0, 1]$ via min-max scaling:
\begin{equation}
    \tilde{\delta}^{(k)}_i = \frac{\delta^{(k)}_i - \delta^{(k)}_{\text{min}}}{ \delta^{(k)}_{\text{max}} -  \delta^{(k)}_{\text{min}}},
\end{equation}
where the min and max are computed across all suffix tokens in the current iteration.
With the normalized $\tilde{\delta}^{(k)}_i$, the aggregated gradient for suffix token $s_i$ can be computed by:
\begin{equation}
    \label{eq:agg_gradient}
    \tilde{g}_i = \sum_{k=0}^{K}\alpha_k \cdot \tilde{\delta}^{(k)}_i \cdot  \frac{\partial \mathcal{L}_A}{\partial \mathbf{z}_{\text{total}}}.
\end{equation}
The gradient for suffix token $s_i$ can be used for adversarial suffix optimization. 



\begin{algorithm}[!t]
\small
\caption{Suffix Optimization Algorithm}
\label{alg:ecgo}
\begin{algorithmic}

\Require
Trained hybrid ensemble surrogate router;
Query set $\mathcal{Q}$ with $|\mathcal{Q}| = m$;
initial suffix $s=[s_1,\dots,s_{L}]$;
loss $\mathcal{L}\coloneqq \mathcal{L}_A$;
iterations $T$;
batch size $B$.
\State $m_c \coloneqq 1$ \Comment{Start by optimising just the first query}
\Loop\ $T$ times
    \For{$i \in [1 \ldots L]$} 
        \State $C_i \coloneqq \mathrm{TopK}(\tilde{g}_i)$ \Comment{Eq.~(9)--(11)}
    \EndFor

    \For{$b = 1,\dots,B$} 
        \State $s^{(b)} \coloneqq s$
        \State Update $s^{(b)}$ by random sample based on its $C_i$.
    \EndFor

    \State $s \coloneqq s^{(b^\star)}$, where
    $b^\star = \arg\min_{b}\, \mathcal{L}\!\bigl(s^{(b)}\bigr)$
    \If{$s$ succeeds on $q_1,\dots,q_{m_c}$ and $m_c < m$}
        \State $m_c \coloneqq m_c + 1$ \Comment{Include the next query}
    \EndIf
\EndLoop
\Ensure Optimized universal suffix $s$
\end{algorithmic}
\end{algorithm}

\noindent \textbf{Suffix Optimization Algorithm}. 
Algorithm~\ref{alg:ecgo} performs adversarial suffix optimization over a single universal suffix $s$ using a hybrid ensemble surrogate router with the aggregated token gradients by Eq.(\ref{eq:agg_gradient}).
At each iteration, the surrogate ensemble produces position-wise scores that define a top-$k$ candidate set $C_i$ for each suffix position. 
We then sample a batch of $B$ variants by replacing a random suffix token with a uniformly sampled candidate from $C_i$, forward them through the ensemble to evaluate $\mathcal{L}_A$, and update $s$ with the variant with the lowest loss. We incorporate new queries incrementally, adding the next one only after the current suffix succeeds on all previously activated queries.

\begin{table*}[!ht]
\centering
\resizebox{\linewidth}{!}{
\small
\setlength{\tabcolsep}{3pt}
\renewcommand{\arraystretch}{1.1} 
\begin{tabular}{llccccccc}
\toprule
& & \multicolumn{3}{c}{In-Distribution Datasets} & \multicolumn{3}{c}{Out-of-Distribution Datasets} & \\
\cmidrule(lr){3-5}\cmidrule(lr){6-8}
\textbf{Target Router} & \textbf{Model} & \textbf{MMLU} & \textbf{GSM8K} & \textbf{MT-Bench} & \textbf{SimpleQA} & \textbf{ArenaHard} & \textbf{RArena} & \textbf{Avg} \\
\midrule

\multirow{6}{*}{\texttt{RouteLLM-Bert}} 
 & clean  & 0.26$_{0.06}$ & 0.50$_{0.02}$ & 0.60$_{0.08}$ & 0.24$_{0.01}$ & 0.55$_{0.04}$ & 0.27$_{0.02}$ & 0.40 \\
 & LifeCycle (W) & 0.45$_{0.09}$ & 0.94$_{0.01}$ & 0.93$_{0.01}$ & 0.58$_{0.04}$ & 0.77$_{0.02}$ & 0.45$_{0.02}$ & 0.69 \\
 & LifeCycle (B) & 0.28$_{0.08}$ & 0.82$_{0.04}$ & 0.72$_{0.03}$ & 0.49$_{0.02}$ & 0.58$_{0.04}$ & 0.31$_{0.02}$ & 0.53 \\
 & Rerouting & 0.58$_{0.01}$ & 0.99$_{0.02}$ & 0.93$_{0.01}$ & 0.81$_{0.02}$ & 0.74$_{0.02}$ & 0.55$_{0.02}$ & 0.77 \\
 & CoT  & 0.32$_{0.05}$ & 0.75$_{0.03}$ & 0.78$_{0.02}$ & 0.38$_{0.03}$ & 0.60$_{0.02}$ & 0.30$_{0.01}$ & 0.52 \\
 & {\method} (Ours) 
 & \textcolor{RubineRed}{0.78$_{0.03}$} (0.52 $\uparrow$)
 & \textcolor{RubineRed}{0.99$_{0.01}$} (0.49 $\uparrow$)
 & \textcolor{RubineRed}{0.93$_{0.01}$} (0.33 $\uparrow$)
 & \textcolor{RubineRed}{1.00$_{0.00}$} (0.76 $\uparrow$)
 & \textcolor{RubineRed}{0.84$_{0.02}$} (0.29 $\uparrow$)
 & \textcolor{RubineRed}{0.82$_{0.02}$} (0.55 $\uparrow$)
 & \textcolor{RubineRed}{0.89}~(0.49 $\uparrow$) \\
\hline

\multirow{6}{*}{\texttt{GraphRouter}} 
 & clean  & 0.50$_{0.10}$ & 1.00$_{0.00}$ & 0.46$_{0.11}$ & 0.69$_{0.02}$ & 0.67$_{0.03}$ & 0.51$_{0.02}$ & 0.64 \\
 & LifeCycle (W) & 0.53$_{0.10}$ & 1.00$_{0.00}$ & 0.46$_{0.11}$ & 0.83$_{0.01}$ & 0.69$_{0.03}$ & 0.63$_{0.04}$ & 0.69 \\
 & LifeCycle (B) & 0.42$_{0.13}$ & 1.00$_{0.00}$ & 0.46$_{0.11}$ & 0.62$_{0.01}$ & 0.63$_{0.03}$ & 0.45$_{0.01}$ & 0.60 \\
 & Rerouting & 0.44$_{0.11}$ & 1.00$_{0.00}$ & 0.46$_{0.11}$ & 0.67$_{0.02}$ & 0.68$_{0.02}$ & 0.50$_{0.01}$ & 0.63 \\
 & CoT & 0.57$_{0.07}$ & 1.00$_{0.00}$ & 0.46$_{0.11}$ & 0.69$_{0.02}$ & 0.62$_{0.03}$ & 0.54$_{0.02}$ & 0.65 \\
 & {\method} (Ours) 
 & \textcolor{RubineRed}{0.84$_{0.03}$} (0.34 $\uparrow$)
 & \textcolor{RubineRed}{1.00$_{0.00}$} (0.00 $\uparrow$)
 & \textcolor{RubineRed}{0.73$_{0.06}$} (0.27 $\uparrow$)
 & \textcolor{RubineRed}{0.94$_{0.01}$} (0.25 $\uparrow$)
 & \textcolor{RubineRed}{0.83$_{0.01}$} (0.16 $\uparrow$)
 & \textcolor{RubineRed}{0.89$_{0.03}$} (0.38 $\uparrow$)
 & \textcolor{RubineRed} {0.87}~(0.23 $\uparrow$) \\
\hline

\multirow{6}{*}{\texttt{P2L}} 
 & clean & 0.74$_{0.02}$ & 0.83$_{0.10}$ & 0.93$_{0.01}$ & 0.16$_{0.01}$ & 0.62$_{0.01}$ & 0.74$_{0.03}$ & 0.67 \\
 & LifeCycle (W) & 0.70$_{0.01}$ & 0.99$_{0.01}$ & 0.90$_{0.03}$ & 0.18$_{0.01}$ & 0.59$_{0.01}$ & 0.63$_{0.03}$ & 0.67 \\
 & LifeCycle (B) & 0.68$_{0.04}$ & 0.98$_{0.02}$ & 0.87$_{0.03}$ & 0.18$_{0.02}$ & 0.63$_{0.01}$ & 0.63$_{0.03}$ & 0.66 \\
 & Rerouting & 0.52$_{0.01}$ & 0.91$_{0.05}$ & 0.83$_{0.02}$ & 0.12$_{0.02}$ & 0.61$_{0.01}$ & 0.52$_{0.05}$ & 0.59 \\
 & CoT & 0.88$_{0.03}$ & 0.97$_{0.04}$ & \textcolor{RubineRed}{0.95$_{0.04}$} & \textcolor{RubineRed}{0.22$_{0.02}$} & 0.62$_{0.02}$ & 0.78$_{0.02}$ & 0.74 \\
 & {\method} (Ours) 
 & \textcolor{RubineRed}{0.89$_{0.03}$} (0.15 $\uparrow$)
 & \textcolor{RubineRed}{1.00$_{0.00}$} (0.17 $\uparrow$)
 & 0.93$_{0.01}$ (0.00 $\uparrow$)
 & 0.18$_{0.02}$ (0.02 $\uparrow$)
 & \textcolor{RubineRed}{0.63$_{0.05}$} (0.01 $\uparrow$)
 & \textcolor{RubineRed}{0.83$_{0.03}$} (0.09 $\uparrow$)
 & \textcolor{RubineRed}{0.74}~(0.07 $\uparrow$) \\
\hline

\multirow{6}{*}{\texttt{RouterDC}} 
 & clean  & 0.83$_{0.00}$ & 0.06$_{0.05}$ & 1.00$_{0.00}$ & 0.68$_{0.02}$ & 0.97$_{0.02}$ & 0.79$_{0.02}$ & 0.72 \\
 & LifeCycle (W) & 0.99$_{0.00}$ & 0.43$_{0.06}$ & 1.00$_{0.00}$ & 1.00$_{0.00}$ & 1.00$_{0.02}$ & 1.00$_{0.00}$ & 0.90 \\
 & LifeCycle (B) & 1.00$_{0.00}$ & 0.46$_{0.09}$ & 1.00$_{0.00}$ & 1.00$_{0.00}$ & 1.00$_{0.02}$ & 1.00$_{0.00}$ & 0.91 \\
 & Rerouting & 0.99$_{0.00}$ & 0.25$_{0.05}$ & 1.00$_{0.00}$ & 1.00$_{0.00}$ & 0.99$_{0.00}$ & 1.00$_{0.00}$ & 0.87 \\
 & CoT  & 0.93$_{0.00}$ & 0.09$_{0.06}$ & 1.00$_{0.00}$ & 0.85$_{0.01}$ & 0.98$_{0.00}$ & 0.89$_{0.03}$ & 0.79 \\
 & {\method} (Ours) 
 & \textcolor{RubineRed}{1.00$_{0.00}$} (0.17 $\uparrow$)
 & \textcolor{RubineRed}{0.61$_{0.09}$} (0.55 $\uparrow$)
 & \textcolor{RubineRed}{1.00$_{0.00}$} (0.00 $\uparrow$)
 & \textcolor{RubineRed}{1.00$_{0.00}$} (0.32 $\uparrow$)
 & \textcolor{RubineRed}{1.00$_{0.02}$} (0.03 $\uparrow$)
 & \textcolor{RubineRed}{1.00$_{0.00}$} (0.21 $\uparrow$)
 & \textcolor{RubineRed}{0.94}~(0.22 $\uparrow$) \\
\hline

\multirow{6}{*}{\texttt{RouteLLM-MF}} 
 & clean  & 0.38$_{0.14}$ & 0.85$_{0.03}$ & 0.27$_{0.05}$ & 0.81$_{0.02}$ & 0.58$_{0.01}$ & 0.44$_{0.03}$ & 0.56 \\
 & LifeCycle (W) & 0.70$_{0.07}$ & 0.99$_{0.01}$ & 0.53$_{0.01}$ & 0.94$_{0.02}$ & 0.71$_{0.02}$ & 0.72$_{0.01}$ & 0.77 \\
 & LifeCycle (B) & 0.45$_{0.13}$ & 0.93$_{0.00}$ & 0.42$_{0.04}$ & 0.85$_{0.02}$ & 0.63$_{0.01}$ & 0.54$_{0.02}$ & 0.64 \\
 & Rerouting  & 0.90$_{0.02}$ & 1.00$_{0.00}$ & 0.65$_{0.06}$ & 1.00$_{0.01}$ & 0.79$_{0.02}$ & 0.93$_{0.01}$ & 0.88 \\
 & CoT  & 0.35$_{0.13}$ & 0.84$_{0.04}$ & 0.33$_{0.03}$ & 0.79$_{ 0.01}$ & 0.54$_{0.03}$ & 0.37$_{0.01}$ & 0.54 \\
 & {\method} (Ours) 
 & \textcolor{RubineRed}{0.98$_{0.01}$} (0.60 $\uparrow$)
 & \textcolor{RubineRed}{1.00$_{0.00}$} (0.15 $\uparrow$)
 & \textcolor{RubineRed}{0.82$_{0.04}$} (0.55 $\uparrow$)
 & \textcolor{RubineRed}{1.00$_{0.00}$} (0.19 $\uparrow$)
 & \textcolor{RubineRed}{0.91$_{0.01}$} (0.33 $\uparrow$)
 & \textcolor{RubineRed}{0.98$_{0.01}$} (0.54 $\uparrow$)
 & \textcolor{RubineRed}{0.95}~(0.39 $\uparrow$) \\
\hline

\multirow{6}{*}{\texttt{OpenRouter}$^{*}$} 
 & clean & 0.12$_{0.17}$ & 0.37$_{0.53}$ & 0.32$_{0.46}$ & 0.00$_{0.00}$ & 0.57$_{0.00}$ & 0.25$_{0.00}$ & 0.27 \\
 & LifeCycle (W)  & 0.35$_{0.00}$ & 0.75$_{0.01}$ & 0.76$_{0.08}$ & 0.04$_{0.05}$ & 0.43$_{0.10}$ & 0.30$_{0.00}$ & 0.44  \\
 & LifeCycle (B)  & 0.34$_{0.01}$ & 0.77$_{0.01}$ & 0.68$_{0.04}$ & 0.00$_{0.00}$ & 0.36$_{0.12}$ & 0.35$_{0.00}$ &0.42 \\
 & Rerouting      & 0.28$_{0.00}$ & \textcolor{RubineRed}{0.91$_{0.05}$} & 0.71$_{0.08}$ & 0.00$_{0.00}$ & 0.54$_{0.15}$ & 0.20$_{0.00}$ & 0.44 \\
 & CoT & 0.24$_{0.01}$ & 0.85$_{0.01}$ & 0.76$_{0.00}$ & 0.00$_{0.00}$ & 0.43$_{0.10}$ & 0.23$_{0.03}$ & 0.42 \\
 & {\method} (Ours)
 & \textcolor{RubineRed}{0.89$_{0.01}$} (0.77 $\uparrow$)
 & 0.88$_{0.01}$ (0.51 $\uparrow$)
 & \textcolor{RubineRed}{0.79$_{0.04}$} (0.47 $\uparrow$)
 & \textcolor{RubineRed}{0.31$_{0.00}$} (0.31 $\uparrow$)
 & \textcolor{RubineRed}{0.61$_{0.15}$} (0.04 $\uparrow$)
 & \textcolor{RubineRed}{0.93$_{0.04}$} (0.68 $\uparrow$) & \textcolor{RubineRed}{0.74}~(0.47 $\uparrow$) \\

\bottomrule
\end{tabular}}
\caption{Average Attack Success Rate and standard deviations are reported for 3 runs. Improvements of our {\method} relative to the clean query, i.e., $s=\emptyset$, are shown. Best results are highlighted in \textcolor{RubineRed}{color}. The target router will be removed from the ensemble pool if an overlap occurs. Out-distribution queries have not been used in either surrogate training or suffix optimization. \texttt{OpenRouter}$^{*}$ is a real-world black-box router.}
\label{tab:main_results}
\end{table*}

\section{Experiments}
In this section, we conduct experiments to answer the following research questions:

\begin{itemize}[leftmargin=*, itemsep=0pt,topsep=0pt]
    \item \textbf{RQ1}: Can {\method} learn an adversarial suffix that effectively directs diverse routers to expensive strong models?
    \item \textbf{RQ2}: Can the learned adversarial suffix be generalized to closed-source routers like GPT-5 and measurably increase inference cost?
    \item \textbf{RQ3}: Is {\method} sensitive to hand-crafted defense mechanisms?
\end{itemize}

\subsection{Experimental Settings}
\noindent \textbf{Target Router}. 
We testify 9 target routers including \texttt{RouteLLM-Bert}, \texttt{GraphRouter}, \texttt{P2L} \texttt{RouterDC}, \texttt{RouteLLM-MF},  \texttt{OpenRouter}, and \texttt{GPT-5}. 
Our ensemble pool consists of five open-source routers, which are listed in Tab.~\ref{tab:router_overview}. 
To strictly separate target and surrogate routers and prevent data leakage, when a target router is included in the ensemble pool, we remove this router from the ensemble pool before the surrogate training.

\noindent \textbf{Datasets}. 
To demonstrate the generalization ability of adversary suffix learned by {\method}, evaluations are conducted on datasets in two settings:
\begin{itemize}[leftmargin=*, itemsep=0pt, topsep=0pt]
    \item  \textbf{In-Distribution}: For surrogate model training and adversarial suffix optimization, we collect three benchmarks, i.e., \textbf{MMLU}, \textbf{GSM8K}, and \textbf{MT-Bench}~\cite{mmliu,gsm8k,mt-bench}. 
    Each dataset is generally split into three disjoint subsets: $\mathcal{D}_{\text{proxy}}$ for surrogate model training, $\mathcal{D}_{\text{suffix}}$ for suffix optimization, and $\mathcal{D}_{\text{eval}}$ for evaluation of in-distribution generalization. The query budget is set to 120 for all experiments. 
    \item \textbf{Out-of-Distribution}: To evaluate the adversarial suffix's generalization on out-of-distribution queries, we apply the suffixes learned from the in-distribution datasets directly to three unseen datasets: \textbf{SimpleQA}, \textbf{ArenaHard}, and \textbf{RArena}~\cite{simpleqa,li2024crowdsourced,lu2025routerarenaopenplatformcomprehensive}.
\end{itemize}
Full dataset statistics are in Appendix~\ref{app:dataset}.

\noindent \textbf{Baselines}. 
The following baselines are compared:
\begin{itemize}[leftmargin=*, itemsep=0pt, topsep=0pt]
    \item \textbf{Rerouting}~\citep{reroutingllmrouters}: A hill-climbing-based attack that discovers query-independent adversarial triggers to maximize the router’s complexity score,  steering queries away from weak models and into strong models.
    \item \textbf{Life-cycle}~\citep{lin2025life}: This paper proposes two universal trigger attacks on LLM routers: \textbf{LifeCycle (W)} and \textbf{LifeCycle (B)}.
    LifeCycle (W) accesses and optimizes a trigger via gradients to maximize strong-model selection.
    LifeCycle (B) extracts a fixed, domain-agnostic trigger from high-win-rate queries via GPT-4o and uses it to induce false-positive routing. 

    \item \textbf{Chain-of-Thought (CoT)}~\citep{kojima2022large}: A simple prompt-engineering baseline that appends \textit{''Let's think step by step''} to inputs, explicitly increasing perceived reasoning complexity to encourage routing to the strong model.
\end{itemize}

\noindent \textbf{Evaluation Metric.}
We evaluate attack effectiveness using the \textbf{Attack Success Rate (ASR)}.
Given a dataset $\mathcal{D}$ and a suffix $s$, ASR measures the fraction of queries routed to the high-capability model set $\mathcal{M}_{\text{strong}}$:
\begin{equation}
    \text{ASR}(s) = \frac{1}{|\mathcal{D}|} \sum_{q \in \mathcal{D}} \mathbb{I}\left( \mathcal{R}_t(q \oplus s) \in \mathcal{M}_{\text{strong}} \right),
\end{equation}
where $\mathcal{R}_t(\cdot)$ denotes the target router and $\mathbb{I}(\cdot)$ is the indicator function.
A higher ASR indicates a more effective adversarial suffix. 

\subsection{Results of Routing Attack}
To answer \textbf{RQ1}, we report the Attack Success Rate (ASR) on six target routers using queries from three in-distribution and three out-of-distribution datasets. Note that out-distribution queries have not been used in either surrogate training or suffix optimization. 
The results in Tab.~\ref{tab:main_results} show:
\begin{itemize}[leftmargin=*, itemsep=0pt, topsep=0pt]
    \item  {\method} consistently achieves state-of-the-art attack success rates (ASRs), substantially outperforming prior adversarial methods. 
    This demonstrates the effectiveness of adversary suffix optimization with a hybrid ensemble router.
    \item Our {\method} maintains high attack success rates across routers and query distributions. This indicates the strong generalization ability of the adversarial suffix learned by {\method}.
\end{itemize}
\begin{figure}[!t]
    \centering
    \includegraphics[width=0.92\linewidth]{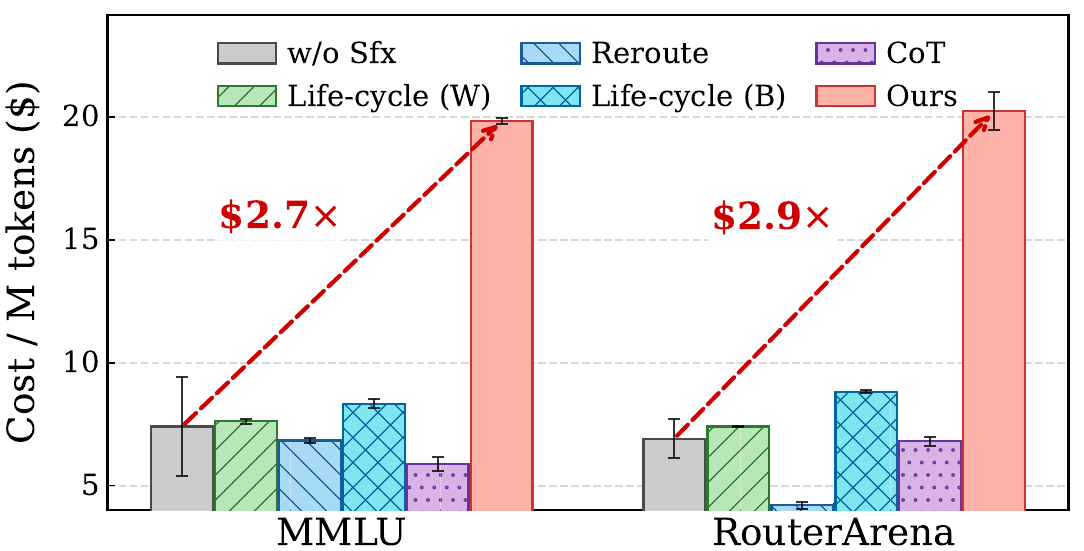}
    \vspace{-0.5em}
    \caption{Inference cost comparisons after attacks.}
    \vskip -1em
    \label{fig:cost_analysis}
\end{figure}

\noindent \textbf{Inference Cost Analysis}. \label{sec:cost_analysis}
To further address \textbf{RQ1}, we analyze the monetary cost reported by the OpenRouter API to assess the economic impact of rerouting, as illustrated in Fig.~\ref{fig:cost_analysis}. 
We find that {\method} leads to a noticeable rise in inference cost. 
On the MMLU benchmark, the average cost per million tokens increases by approximately $\mathbf{2.7\times}$ compared to the clean baseline. 
The effect is slightly stronger on the out-of-distribution dataset RouterArena, with a $\mathbf{2.9\times}$ increase. 
These results indicate that the router is frequently redirected to higher-cost models under our optimized suffixes. 
On the adversarial side, the cost of mounting this attack is remarkably low. Collecting the 120 surrogate training queries requires a total investment of only \$0.98 . While the suffix-induced variation in completion length is dataset-dependent, the overall financial overhead remains negligible. These costs are detailed in ~\ref{app:Attacker cost}.

\begin{figure}[t]
    \centering
    
        \includegraphics[width=\linewidth]{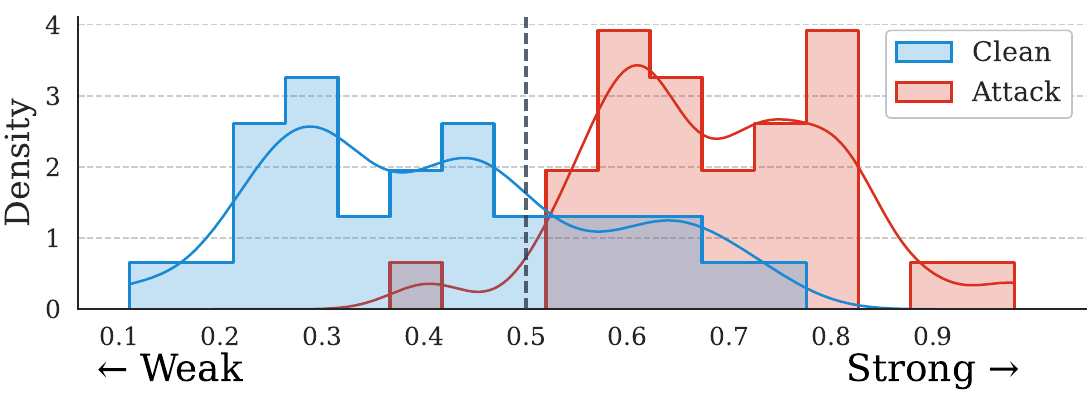}
        \caption{Distribution of fingerprinting scores with strong \texttt{GPT-5-Thinking} models. After attacked by {\method}, responses are more likely from strong models.}
    
    \label{fig:GPT_Bow_POS_probs}
\end{figure}
\begin{table}[!t]
    \centering
    \small
    \begin{tabularx}{\columnwidth}{l *{2}{>{\centering\arraybackslash}X}}
        \toprule
        \textbf{Metric} & \textbf{Clean} & \textbf{Attack} \\
        \midrule
        Comprehensiveness & 36.0\%  & \textbf{64.0\% } \\
        Diversity         & 28.0\% & \textbf{72.0\% } \\
        Empowerment       & 36.0\%  & \textbf{64.0\% } \\
        Overall  & 36.0\%  & \textbf{64.0\% }\\
        \bottomrule
    \end{tabularx}
    \caption{Win rates (\%) of clean queries v.s. attacked queries across four evaluation dimensions.}
    \label{tab:quality}
\end{table}

\subsection{Attacking GPT-5 Router}
\label{subsec:gpt5_web_study}
To address \textbf{RQ2}, we conduct a study on the web-based GPT-5 interface to evaluate whether {\method} generalizes to closed-source commercial routers, whose routing decisions are unknown.

\noindent\textbf{Setup of Attacking GPT-5 Router.}
The GPT-5 web interface provides three modes \texttt{Auto}, \texttt{Instant}, and \texttt{Thinking}, which implicitly trade off cost and latency. 
As GPT-5 exposes no routing decisions, we directly apply adversarial suffixes trained on OpenRouter. 
We randomly sample 50 questions from the OOD test set and query GPT-5 in \textit{Auto} mode with and without the attack suffix. 
All interactions are conducted in temporary sessions to avoid personalization effects.

\noindent\textbf{Evaluation of on GPT-5 Router.} 
For the GPT-5 router, ASR can not be computed due to the lack of routing decisions. 
Instead, we evaluate the effectiveness of {\method} indirectly from two aspects: 
\begin{itemize}[leftmargin=*, itemsep=0pt, topsep=1pt]
    \item \textbf{Response Quality}: We first test the impact of the suffixes trained on \texttt{RouteLLM-BERT} and \texttt{RouteLLM-MF} by testing them on a fixed \texttt{GPT-4} backend. Table~\ref{tab:fixed_backend_quality} show no performance drop on GSM8K, suggesting that the suffixes do not degrade generation quality. Given these results, we evaluate the GPT-5 router using an LLM judge to compare responses with and without adversarial suffixes, following~\citet{guo-etal-2025-lightrag}. Since routing to stronger models should yield better answers even for simple tasks, a higher win rate for the latter indicates that \method{} effectively misleads the router to more expensive models.
    \item \textbf{Fingerprting Score with Strong Model}: We infer the routing decision via Bag-of-Words fingerprinting~\citep{bai-etal-2025-esf,mcgovern-etal-2025-large,yan2025duffin}. Specifically, we treat responses generated in \texttt{Thinking} mode as a proxy for the strong model style. The resulting {Thinking}-likeness score is interpreted as the probability that a query is routed to the strong model.
\end{itemize}

\begin{figure}[t]
    \centering
    \begin{tcolorbox}[
        colback=gray!2, 
        colframe=black!85, 
        title=\textbf{Case Study: Rerouting GPT-5},
        boxrule=1pt,
        arc=2pt,
        fonttitle=\bfseries
    ]
    \small
    \textbf{User Query:} "What is the pressure of carbon dioxide at 200°F and a specific volume of 0.20 $ft^3/lbm$?"
    
    \vspace{0.4em}
    \hrule
    \vspace{0.6em}

\begin{minipage}[c]{0.45\linewidth}
        \textbf{Clean Prompt} \\
        \textit{Input:} User Query
    \end{minipage}
    \hfill
    \begin{minipage}[c]{0.52\linewidth}
        \begin{tcolorbox}[
            colback=white, colframe=black!35, 
            boxrule=0.5pt, arc=1pt,
            left=3pt, right=3pt, top=3pt, bottom=3pt,
            boxsep=0pt
        ]
            \scriptsize \textbf{\faClock \ Internal} \hfill \colorbox{gray!15}{\textbf{Duration: 0s}} \\
            \textit{"None"}
        \end{tcolorbox}
    \end{minipage}

    \vspace{0.3em}
    \textit{Output:}  \dots Result: $P \approx 8.0 \times 10^2$ psi. \textcolor{red}{\faTimes}
    \centerline{\hdashrule{0.95\linewidth}{0.5pt}{2pt 2pt}} 
    
    \textbf{{\method}} \\
    \textit{Input:} User Query + \underline{\texttt{[Universal Suffix]}} 
    
    \vspace{0.15em} 
    \begin{tcolorbox}[
            colback=red!5,         %
            colframe=red!30,       %
            boxrule=0.6pt, 
            arc=1pt, 
            left=4pt, right=4pt, top=4pt, bottom=4pt
        ]
        \scriptsize \color{black!80!black} 
        \textbf{\faClock \ Internal Thought Process} \hfill \colorbox{green!15}{\textbf{Duration: 51s}} \\
        \vspace{0.2em}
        \begin{tabular}{@{}p{0.95\linewidth}@{}}
            \textit{1. Calculating thermodynamic properties of CO\textsubscript{2}} \\
            \textit{2. Evaluating CO\textsubscript{2} properties and resolving terminology} \\
            \textit{3. Solving PR equation for pressure calculation} \\
            \textit{4. Performing temperature and volume conversions} \\
            \textit{[Reasoning continues] \dots}
        \end{tabular}
    \end{tcolorbox}
    
    \textit{Output:} ...the estimate is: {$P \approx 695$ psia}. \textcolor{ForestGreen}{\faCheck}
    
    \end{tcolorbox}
    \caption{Case study of on GPT-5: the router switches from a brief incorrect answer (top) to a multi-step reasoning process that yields the correct answer (bottom). This implies that adversarial suffix manage direct GPT-5 router to a stronger model.}

    \label{fig:attack_case_study}
    \vspace{-1em}
\end{figure}
\noindent\textbf{Results of Attacking GPT-5 Router.}
As shown in Fig.~\ref{fig:GPT_Bow_POS_probs}, attacked queries exhibit a clear shift toward higher \texttt{Thinking}-likeness probabilities compared to clean queries. 
From Tab.~\ref{tab:quality}, we observe that attacked responses consistently outperform clean responses across all evaluation dimensions, further confirming the effectiveness of \method.
We also conduct a case study on the GPT-5 \texttt{Auto} interface, which is presented in Fig.~\ref{fig:attack_case_study}. 
We can observe that with the adversarial suffix from {\method}, the router switches from a brief incorrect answer to a multi-step reasoning process that yields the correct answer. The processing time also increases significantly. 
The above observations indicate that {\method} reliably increases the likelihood of routing into the more expensive mode in GPT-5 series.

\subsection{Whitespace Defense}
To address \textbf{RQ3}, we use the whitespace defense that inserts spaces into the suffix~\citep{robeysmoothllm} as a representative example and evaluate our Triggering suffix on three target routers across two datasets. 
As shown in Tab.~\ref{tab:defense_robustness}, {\method} shows a slight decrease in success rate on all three datasets, indicating that it has resistance to specific defense.

\begin{table}[ht!]
    \centering
    \resizebox{\linewidth}{!}{
        \begin{tabular}{cccc}
            \toprule
             & \textbf{RouteLLM-BERT} & \textbf{Graph-Router} & \textbf{RouterDC} \\
            \midrule
            MT-Bench   & 0.95 (0.93) & 0.71 $\downarrow$ (0.73) & 1.00 (1.00) \\
            ArenaHard  & 0.81 $\downarrow$ (0.84) & 0.73 $\downarrow$ (0.83) & 1.00 (1.00)  \\
            \bottomrule
        \end{tabular}
    }
    \caption{Robustness of {\method} under whitespace defense.}
    \label{tab:defense_robustness}
\end{table}

\subsection{Ablation Study}
We ablate two core components of $\text{R}^{2}\text{A}$: LoRA-based surrogate training and gradient normalization, with results shown in Tab.~\ref{tab:two_module}. 
Removing gradient normalization causes consistent performance drops, particularly on MF (0.95 → 0.49), underscoring its role in handling heterogeneous gradient scales. 
Disabling the lightweight router degrades performance, most notably on RouterDC (0.83 → 0.30), indicating the importance of parameter-efficient adaptation under limited queries. 
Overall, both components are necessary for robust and transferable attacks across routers.

\begin{table}[!t]
  \centering
\setlength{\tabcolsep}{2pt}
  \resizebox{\linewidth}{!}{
  \begin{tabular}{lccll}
    \toprule
    \textbf{Model} & 
    \textbf{RouterDC} &
    \textbf{CausalLLM} &
    \textbf{MF} &
    \textbf{SW} \\
    \midrule
    $\text{R}^{2}\text{A}$ & \textbf{0.83} & \textbf{0.83} & \textbf{0.95} & \textbf{0.81} \\
    w/o Lightweight Router & 0.30 & 0.75 & 0.70 & 0.61 \\
    w/o Grad Norm & 0.33 & 0.78 & 0.49 & 0.63 \\
    \bottomrule
  \end{tabular}
  }
  \caption{Ablation studies across in-distribution datasets.}
  \label{tab:two_module}
\end{table}

\begin{figure}[!t]
    \centering
    \includegraphics[width=1.00\linewidth]{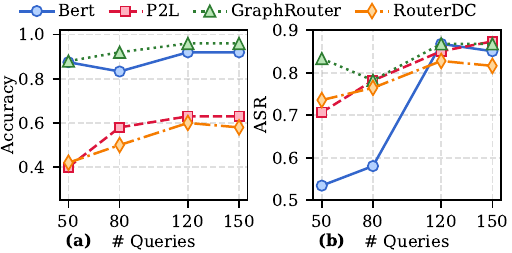}
    \caption{Performance analysis showing Accuracy (a) and ASR (b) trends with varying query counts.}
    \label{fig:query_number}
\end{figure}
\subsection{Impacts of the Query Budget}
We study the effect of surrogate training set size, varying it from 50 to 150 queries, on attack performance.
As shown in Fig.~\ref{fig:query_number}, increasing the query budget consistently improves surrogate accuracy, measured as agreement with the target router’s routing decisions. 
Higher surrogate accuracy leads to higher ASR in turn, indicating a strong correlation between surrogate fidelity and attack effectiveness. 
In particular, this trend is evident for the RouteLLM-Bert, where ASR rises sharply from 0.58 to 0.87 as the budget increases from 80 to 120 queries. 
For most target routers, performance saturates at ~120 queries, with marginal gains thereafter.
This suggests that $\text{R}^{2}\text{A}$ is highly sample-efficient, requiring only a modest number of queries to achieve strong attack performance across heterogeneous routers.
\section{Related Work}
\noindent \textbf{LLM Routers.}
To select the most suitable LLM for a given prompt, a range of routing methods has been proposed.
Early work \citep{chen2023frugalgptuselargelanguage,llm-blender-2023,aggarwal2025automixautomaticallymixinglanguage,zhang2025beyond} focuses on querying multiple LLMs for a single input to select the best response, whereas later approaches \citep{ding2024hybrid,routellm,zooter} aim to predict the best model before the inference stage using different data sources and backbone models.
More recent work improves the ability to capture differences between models through several strategies, including dual contrastive learning \citep{chen2024routerdc}, graph-based learning \citep{feng2024graphrouter}, compact model embeddings \citep{zhuang2025embedllm}, and ranking-based methods such as Elo ratings \citep{Eagle} and Bradley--Terry models \citep{frick2025prompttoleaderboard}, as well as in-context-learning based routers \citep{iclrouter}.
In addition, several routing benchmarks have been introduced to train and evaluate LLM routers \citep{hu2024routerbench,huang2025routereval,feng2025fusionfactory,lu2025routerarenaopenplatformcomprehensive}.

\noindent \textbf{Router Attacks.}
Despite their cost–performance benefits, recent work has exposed vulnerabilities in LLM routers.
\citet{kassem2025robust} show that many routers rely on category-based heuristics, introducing safety risks, and \citet{huang2025exploring} demonstrate that voting-based leaderboards such as Chatbot Arena are vulnerable to adversarial vote manipulation.Closest to our setting, \citet{reroutingllmrouters} and \citet{lin2025life} perturb queries to change routing decisions, but they either assume access to router parameters and gradients or depend on fixed optimization prompts.
In contrast, we optimize suffixes against each target router in a strict black-box setting, using only its observed routing decisions.

\section{Conclusion}
In this paper, we propose a black-box routing attack that learns a universal suffix to reroute LLM routers. 
Using a hybrid ensemble surrogate and an encoder-consistent objective, we optimize a target-specific suffix that biases routing towards stronger and more expensive models. 
Experiments on 7 routers and 6 datasets, including real-world evaluation, show strong effectiveness and generalization ability.
These findings position routing as a security-critical boundary and motivate future stronger monitoring for cost-aware routers.

\section{Limitations}
In this work, we study a black-box optimization attack on LLM routing and train a separate adversarial suffix for each target router to reroute simple queries from cheap to expensive models. 
This study has two main limitations. 
First, we mainly focus on steering queries towards a stronger and typically more expensive model, whereas in practice, some users may wish to target a specific model for other reasons, such as latency or safety, which we do not systematically investigate.
Second, the attack assumes access to the router’s candidate model list and to the identity of the model selected for each query, an assumption that may not hold in some deployments. 

\section{Acknowledgment}
This material is based upon work supported by, or in part by, the National Natural Science Foundation of China (NSFC) under Grant No. 62506316, and the Guangdong Provincial Program under Grant No. 2025DO3JOO15. The findings in this paper do not necessarily reflect the views of the funding agencies.
\nocite{stealing}
\nocite{learning}
\nocite{unizyme}

\bibliography{custom}
\clearpage
\appendix
\raggedbottom

\section{Dataset}\label{app:dataset}
\subsection{Dataset Information}
To ensure a comprehensive evaluation across conversational, knowledge-intensive, and reasoning capabilities, we use three standard benchmarks as our primary training sources and then test on all six datasets listed below:

\begin{itemize}[leftmargin=*]
\item \textbf{MT-Bench-101}~\citep{mt-bench}: A multi-turn conversational benchmark for assessing instruction following and coherence in complex dialogue settings.
\item \textbf{MMLU}~\citep{mmliu}: A large-scale multitask benchmark spanning 57 subjects across STEM, humanities, and social sciences, serving as a proxy for broad world knowledge.
\item \textbf{GSM8K}~\citep{gsm8k}: A collection of high-quality grade-school mathematics problems designed to evaluate multi-step reasoning and logical consistency.
\end{itemize}

These three datasets form the basis of our training and in-domain evaluation.  
To study generalization beyond the training distribution, we further include three evaluation-only benchmarks that are never used during surrogate training or suffix optimization:

\begin{itemize}[leftmargin=*]
\item \textbf{SimpleQA}~\citep{simpleqa}: A short-form question answering benchmark targeting factual correctness on long-tail knowledge.
\item \textbf{Arena Hard}~\citep{li2024crowdsourced}: A set of challenging real-world queries for evaluating model helpfulness and preference alignment.
\item \textbf{RouterArena}~\citep{lu2025routerarenaopenplatformcomprehensive}: A benchmark for evaluating LLM routing across diverse tasks.
\end{itemize}

\subsection{Dataset Split}
We partition the data into three disjoint sets to reflect a realistic attack scenario.

(1) Surrogate router training set $\mathcal{D}_{\text{proxy}}$: 
To model a resource-constrained attacker, we sample a minimal set of 120 queries, consisting of 40 balanced samples from each of the three primary benchmarks (MT-Bench-101, MMLU, GSM8K). This set is used exclusively to train the surrogate router ensemble and to align the projection layers. 

(2) Suffix optimization set $\mathcal{D}_{\text{suffix}}$: 
We sample a separate set of 600 queries (200 per primary benchmark). A 70\% random split (420 queries) is used to perform gradient-based optimization for adversarial suffix generation, while the remaining 30\% (180 queries) is reserved for in-domain evaluation.

(3) Evaluation set $\mathcal{D}_{\text{eval}}$: 
Attack performance is assessed on both in-domain and out-of-domain data.  
The in-domain test set corresponds to the remaining 180 queries from $\mathcal{D}_{\text{suffix}}$. 
For out-of-distribution evaluation, we construct three held-out pools: 500 queries from SimpleQA, 750 from Arena Hard (full set), and 809 from RouterArena. 
All three pools are disjoint from $\mathcal{D}_{\text{proxy}}$ and $\mathcal{D}_{\text{suffix}}$. 
At test time, we uniformly sample 70\% of each pool as the evaluation set and report performance on these prompts, which are never seen during surrogate training or suffix optimization.

\begin{table}[t]
\centering
\small
\setlength{\tabcolsep}{8pt}
\renewcommand{\arraystretch}{1.05}
\begin{tabular}{l c}
\toprule
\textbf{Hyperparameter} & \textbf{Value} \\
\midrule
Adapter rank ($r$) & 16 \\
Epochs & 20 \\
Learning rate & 0.03 \\
Batch size & 32 \\
Optimizer & AdamW \\
\bottomrule
\end{tabular}
\caption{Hyperparameters for hybrid ensemble surrogate router training.}
\label{tab:hyper_ensemble}

\end{table}

\begin{table}[t]
\centering
\small
\setlength{\tabcolsep}{8pt}
\renewcommand{\arraystretch}{1.05}
\begin{tabular}{l c}
\toprule
\textbf{Hyperparameter} & \textbf{Value} \\
\midrule
Optimization iterations ($T$) & 3000 \\
Candidate batch size ($B$) & 64 \\
Top-$k$ sampling & 256 \\
The limit of suffix tokens ($\Delta$)& 30\\
Initial suffix & \texttt{! ! ! ! ! ! ! ! ! !} \\

\bottomrule
\end{tabular}
\caption{Hyperparameters for adversarial suffix optimization.}
\label{tab:hyper_suffix}

\end{table}

\begin{table*}[t]
    \centering
    \small %

    \label{tab:model_pools}
    \newcolumntype{M}{>{\centering\arraybackslash}X} 
    \newcolumntype{L}{>{\raggedright\arraybackslash}X}     \newcolumntype{C}[1]{>{\centering\arraybackslash}m{#1}}

    \begin{threeparttable}
        \begin{tabularx}{\textwidth}{C{90pt} XX}
        
            \toprule
            \textbf{Router} & \textbf{Strong Model Pool} & \textbf{Weak Model Pool} \\
            \midrule
            
            RouterLLM-MF & 
            gpt-4-1106-preview & 
            mixtral-8x7B \\ 
            \addlinespace
            
            RouterLLM-BERT & 
            gpt-4-1106-preview & 
            mixtral-8x7B \\ 
            \addlinespace
            
            RouterLLM-SW & 
            gpt-4-1106-preview & 
            mixtral-8x7B \\ 
            \addlinespace
            
            RouterLLM-CLM & 
            gpt-4-1106-preview & 
            mixtral-8x7B \\ 
            \addlinespace
            
            P2L\tnote{*} & 
            gemini-1.5-pro-exp-0801; gemini-2.0-flash-lite-preview-02-05; gemini-exp-1114; gemini-exp-1121; gemini-exp-1206; glm-4-plus-0111; gemini-1.5-pro-002; deepseek-r1;deepseek-v3; o1-2024-12-17; o1-mini; o1-preview; o3-mini; o3-mini-high; qwen-plus-0125; qwen2.5-max; gemini-1.5-pro-exp-0827; gemini-2.0-flash-001;
            chatgpt-4o-latest-20240808;chatgpt-4o-latest-20240903; chatgpt-4o-latest-20241120;  & 
            rwkb-4-raven-14B; gemma-2b-it; amazon-nova-lite-v1.0; jamba-1.5-mini;  athene-70b-0725; ;llama-3.2-3b-instruct; zephyr-7b-alpha; c4ai-aya-expanse-32b; c4ai-aya-expanse-8b; yi-lightning-lite; mpt-7b-chat; granite-3.0-2b-instruct; gpt-3.5-turbo-0613; gpt-3.5-turbo-1106; llama-3-8b-instruct; llama-3.1-tulu-3-8b; llama-3.2-1b-instruct; oasst-pythia-12b; openchat-3.5;amazon-nova-pro-v1.0 \dots\\
            \addlinespace
            \addlinespace

            GraphRouter & 
            lama-3.1-turbo-70b; llama-3-turbo-70b; qwen-1.5-72b; llama-3-70b; mixtral-8x7b & 
            llama-3-turbo-8b; llama-3-7b; llama-2-7b; mistral-7b; nousresearch \\
            \addlinespace

            RouterDC &
            dolphin2.9-llama-3-8b; dolphin2.6-mistral-7b & 
            metamath-mistral-7b; chinese-mistral-7b; zephyr-7b-beta; llama-3-8b; mistral-7b \\
            \addlinespace

            OpenRouter\tnote{*} &
            claude-opus-4.1; gpt-5; gemini-2.5-pro; 
            claude-opus-4.5; gpt-5.1; gemini-3-pro & 
            mixtral-8x7b-instruct; perplexity-sonar;  
            qwen3-14b; llama-3.1-8b-instruct \dots\\
            
            \bottomrule
        \end{tabularx}
        
    \end{threeparttable}
    \caption{Strong and weak model pool partitions for each router. Full model list of P2L is available at: \url{https://huggingface.co/lmarena-ai/p2l-7b-grk-02222025/blob/main/model_list.json}. Full model list of OpenRouter is available at: \url{https://openrouter.ai/openrouter/auto}. }
    \label{tab:model_partion}
\end{table*}
\begin{table}[!ht]

    \centering

    \small
    \setlength{\tabcolsep}{2pt}
    \begin{tabular}{@{}l l l@{}}
        \toprule
        Router          & Encoder      & Routing Mechanism \\
        \midrule
        RouteLLM-BERT   & XLM-R-base   & Classification \\
        RouteLLM-Causal & Llama-3-8B   & Next-token routing \\
        P2L             & Qwen2.5-7B   & Bradley--Terry ranking \\
        GraphRouter     & MiniLM-L    & Graph-based routing \\
        RouterDC        & mDeBERTa-v3  & Dual contrastive routing \\
        \bottomrule
    \end{tabular}
    \caption{Open-source routers used in surrogate ensemble, with their encoders and routing mechanisms.}
\label{tab:router_overview}
\end{table}
\section{Implementation Details}
\label{sec:appendix_implementation}

\noindent \textbf{Experimental Environment}.
We train our surrogate router and optimize the adversarial suffixes using a high-performance computing cluster. All experiments are conducted on a server equipped with 8 NVIDIA RTX A6000 GPUs and 512\,GB system memory.

\noindent \textbf{Detailed Hyperparameters}. 
The training of the Hybrid Ensemble Surrogate Router and the execution of the ECGO algorithm involve several key hyperparameters. These are detailed in Table \ref{tab:hyper_ensemble} and Table \ref{tab:hyper_suffix}.

\noindent \textbf{Efficiency and Runtime}. 
On a single NVIDIA RTX A6000 GPU, one ECGO iteration takes approximately 31.5 seconds. 
We set $T{=}3000$ as an upper bound on the number of iterations. 
In practice, optimization typically terminates earlier via early stopping once the suffix achieves the target success criterion on the training set, resulting in substantially shorter runtimes. 
After a suffix is obtained, applying it at inference time adds negligible overhead.

\section{Routers and Model Pools}
\subsection{Observability of Commercial Routing Decisions}
\label{app:router_observability}

To justify our black-box assumption, we survey several prominent commercial routing platforms. As shown in Table~\ref{tab:router_transparency}, these services typically operate with high transparency to maintain accountability. They explicitly list their supported model pools and include the final routing decision in the API response metadata, confirming that our threat model aligns with real-world deployments.

\newcommand{\greencheck}{{\color{ForestGreen}\checkmark}}

\begin{table}[!ht]
  \centering
  \small 
  \setlength{\tabcolsep}{15pt} 
  \renewcommand{\arraystretch}{1.3}
  
  \begin{tabular}{lcc}
    \toprule
    \makecell[l]{\textbf{Platform}} &
    \makecell[c]{\textbf{Exposed}\\\textbf{Pool}} &
    \makecell[c]{\textbf{Observable}\\\textbf{Decision}} \\
    \midrule
    OpenRouter\textsuperscript{\ref{fn:openrouter}} & \greencheck & \greencheck \\  
    
    Switchpoint\tablefootnote{\url{https://www.switchpoint.dev/}} & \greencheck & \greencheck \\
    NotDiamond\tablefootnote{\url{https://docs.notdiamond.ai/reference/list_models_v2_models_get}} & \greencheck & \greencheck \\
    Azure-Router\tablefootnote{\url{https://ai.azure.com/catalog/models/model-router}} & \greencheck & \greencheck \\
    \bottomrule
  \end{tabular}
  
  \caption{Overview of model pool visibility and routing decision observability. \greencheck~indicates features are supported and visible to users.}
  \label{tab:router_transparency}
\end{table}
\begin{table*}[t]
\centering
\resizebox{\linewidth}{!}{
\small
\setlength{\tabcolsep}{3pt}
\renewcommand{\arraystretch}{0.95} 
\begin{tabular}{llccccccc}
\toprule
\textbf{Target Router} & \textbf{Model} & \textbf{MMLU} & \textbf{GSM8K} & \textbf{MT-Bench} & \textbf{SimpleQA} & \textbf{ArenaHard} & \textbf{RArena} & \textbf{Avg} \\
\midrule

\multirow{5}{*}{\texttt{RouteLLM-CLM}} 
 & clean  & 0.38$_{0.04}$ & 0.99$_{0.01}$ & 0.43$_{0.07}$ & 0.97$_{0.02}$ & 0.58$_{0.02}$ & 0.36$_{0.02}$ & 0.62 \\
 & LifeCycle (B) & 0.69$_{0.05}$ & 1.00$_{0.00}$ & 0.66$_{0.07}$ & 1.00$_{0.00}$ & 0.79$_{0.02}$ & 0.66$_{0.03}$ & 0.80 \\
 & CoT  & 0.41$_{0.03}$ & 0.99$_{0.01}$ & 0.45$_{0.09}$ & 0.98$_{0.00}$ & 0.58$_{0.04}$ & 0.38$_{0.02}$ & 0.63 \\
 & {\method} (Ours) 
 & {0.71$_{0.05}$} (0.33 $\uparrow$)
 & {1.00$_{0.00}$} (0.01 $\uparrow$)
 & 0.60$_{0.08}$ (0.17 $\uparrow$)
 & {1.00$_{0.03}$} (0.03 $\uparrow$)
 & {0.82$_{0.00}$} (0.24 $\uparrow$)
 & {0.70$_{0.02}$} (0.34 $\uparrow$)
 & 0.81 (0.19 $\uparrow$) \\
\hline

\multirow{5}{*}{\texttt{RouteLLM-SW}} 
 & clean  & 0.13$_{0.03}$ & 0.31$_{0.03}$ & 0.33$_{0.03}$ & 0.03$_{0.01}$ & 0.44$_{0.03}$ & 0.21$_{0.01}$ & 0.24 \\
 & LifeCycle (B) & 0.59$_{0.03}$ & 0.80$_{0.04}$ & 0.48$_{0.03}$ & 0.63$_{0.04}$ & 0.73$_{0.02}$ & 0.62$_{0.00}$ & 0.64 \\
 & CoT  & 0.21$_{0.01}$ & 0.60$_{0.04}$ & 0.40$_{0.02}$ & 0.12$_{0.02}$ & 0.55$_{0.03}$ & 0.30$_{0.01}$ & 0.36 \\
 & {\method} (Ours) 
 & {0.75$_{0.01}$} (0.62 $\uparrow$)
 & {0.93$_{0.03}$} (0.62 $\uparrow$)
 & {0.50$_{0.05}$} (0.17 $\uparrow$)
 & {0.77$_{0.04}$} (0.74 $\uparrow$)
 & {0.86$_{0.01}$} (0.42 $\uparrow$)
 & {0.79$_{0.02}$} (0.58 $\uparrow$)
 & 0.77 (0.53 $\uparrow$)\\
\bottomrule
\end{tabular}}
\caption{Supplemental results for RouteLLM-CLM and RouteLLM-SW.}

\label{tab:supp_results}

\end{table*}
\subsection{Surrogate Ensemble of Routers}

We construct a surrogate ensemble from five heterogeneous open-source LLM routers: RouteLLM-Bert, RouteLLM-Causal, P2L, GraphRouter, and RouterDC, as summarized in Table~\ref{tab:router_overview}.
These routers span diverse backbone encoders, including encoder-only models, causal language models, and sentence encoders, and employ distinct routing mechanisms such as supervised classification, Bradley--Terry ranking, graph-based routing, and dual contrastive objectives.
Their candidate model pools also differ in both size and composition.
In our setting, we do not reproduce the original deployment configuration of each router; rather, we treat this heterogeneous collection as a unified surrogate ensemble that supplies gradients for adversarial suffix optimization.

For all routers, we prioritize the use of publicly available weights and follow the default configurations and datasets provided in their official repositories.For the trainable lightweight router  introduced in the main text, we instantiate its encoder with the public \texttt{all-MiniLM-L6-v2}~\cite{MINILM}. 
All routing models are trained and evaluated on a cluster with eight NVIDIA RTX A6000 GPUs, which ensures a consistent computational environment across experiments.

\subsection{Target routers and evaluation setting}

During evaluation, the five routers that constitute the surrogate ensemble in Table~\ref{tab:router_overview} also serve as target routers. 
In addition, we treat RouteLLM-MF, RouteLLM-SW, and the commercial black-box aggregator OpenRouter as further target routers. 
For a given target router $R$, we remove $R$ from the surrogate ensemble and exclude its outputs from both the surrogate loss and the gradient computation. 
Adversarial suffixes are optimized only with respect to the remaining surrogate routers and are then applied to the held-out target router.
\subsection{Strong vs.\ Weak Model Partition}
\label{app:Model Partition}

For each router, we follow its original candidate model pool and split the models into a strong tier $\mathcal{M}_{\text{strong}}$ and a weak tier $\mathcal{M}_{\text{weak}}$, as summarized in Table~\ref{tab:model_partion}. 
The split is determined using public leaderboards, model size, and provider-side cost, so that higher-capacity and more expensive models are assigned to $\mathcal{M}_{\text{strong}}$ and lighter, cheaper models to $\mathcal{M}_{\text{weak}}$. During suffix optimization, we maximize the probability, as estimated by the surrogate ensemble, that a query is routed into $\mathcal{M}_{\text{strong}}$. 
Our main Attack Success Rate (ASR) metric then measures how often the target router’s decision changes from a weak to a strong model after appending the universal suffix. 
This two-tier structure of $\mathcal{M}_{\text{strong}}$ and $\mathcal{M}_{\text{weak}}$ underpins the attack formulation.

\definecolor{cotblue}{RGB}{218, 232, 252}   
\definecolor{cotgreen}{RGB}{213, 232, 212}  
\definecolor{cotred}{RGB}{248, 206, 204}    
\definecolor{cotgray}{RGB}{245, 245, 245}   

\section{Additional Results }

\subsection{Supplementary Results}
For a fair comparison, we apply all triggers as suffixes in our experiments.
For baselines that require router parameters and gradients during optimization, we first optimize a universal trigger on a selected source router following the original setup, and then evaluate it on other target routers via transfer, without any target-side optimization.
All evaluations on the target routers are conducted in a black-box setting. 

We report additional results on RouteLLM-CLM and RouteLLM-SW in Table~\ref{tab:supp_results}. Note that the Rerouting attack is optimized on RouteLLM‑SW  while LifeCycle(W) is trained on RouteLLM‑CLM. As these two models are white‑box to their corresponding attack methods, we omit those results from the table.



\subsection{Impact of Suffixes on Generation Quality}
\label{app:quality_check}

To isolate the impact of the adversarial suffix from the effects of model switching, we conducted a controlled experiment using a fixed \texttt{GPT-4} backend. We randomly sampled 30 questions from the GSM8K dataset and compared the model's accuracy with and without the learned suffixes derived from \texttt{RouteLLM-MF} and \texttt{RouteLLM-BERT}. 

As shown in Table~\ref{tab:fixed_backend_quality}, the exact-match accuracy did not suffer any degradation upon appending the suffixes.These results confirm that the performance gains observed in our main GPT-5 experiments are driven by successful model redirection.
\begin{table}[!ht]
    \centering
    \small
    \setlength{\tabcolsep}{10pt}
    \renewcommand{\arraystretch}{1.1}
    \begin{tabular}{lcc}
        \toprule
        \textbf{Suffix Source} & \textbf{With Suffix} & \textbf{Without Suffix} \\
        \midrule
        RouteLLM-MF & \textbf{87.8\%} & 81.6\% \\
        RouteLLM-Bert & \textbf{89.7\%} & 89.7\% \\
        \bottomrule
    \end{tabular}
    \caption{Fixed-backend accuracy check on 30 GSM8K questions using GPT-4.}
    \label{tab:fixed_backend_quality}
\end{table}
\subsection{Cost and Token Overhead}
\label{app:Attacker cost}
We analyze the practical cost of {\method} in terms of the training query budget and the additional token overhead, both quantified using OpenRouter logs. For surrogate training, we use a fixed budget of 120 queries with a total cost of \$0.9826 ($\approx$ \$0.00819 per query). For suffix overhead, Table~\ref{tab:attack_cost} reports the change in average completion length per query.
These results show that the suffix overhead varies across datasets. Notably, the overhead can be negative (e.g., in SimpleQA) when rerouting yields more concise completions.

\begin{table}[!h]
    \centering
    \small
    \setlength{\tabcolsep}{12pt}
    \renewcommand{\arraystretch}{1.1}
    \begin{tabular}{lrr}
        \toprule
        \textbf{Dataset} & \textbf{Clean} & \textbf{Attack} \\
        \midrule
        ArenaHard   & 2006.3 & 1885.7 \\
        SimpleQA    & 1034.0 & 513.3  \\
        RouterArena & 356.1  & 740.0  \\
        \bottomrule
    \end{tabular}
    \caption{Comparison of average completion tokens per query between Clean and Attack states.}
    \label{tab:attack_cost}
\end{table}

\subsection{Thinking-likeness Classifier}
To quantify the “Thinking’’ fingerprint of a reply, we train a lightweight bag-of-words logistic-regression classifier on GPT-5 outputs labeled as \emph{Thinking} versus \emph{Instant}. Each reply is represented with TF–IDF word $n$-grams ($n=1\text{–}3$), capped at 400{,}000 features, and we optimize an $\ell_2$-regularized logistic regression with regularization parameter $C = 30.0$ and a maximum of 4{,}000 iterations. The fingerprint score is given by the classifier’s predicted probability of the \emph{Thinking} class; Table~\ref{tab:style_likeness} reports summary statistics.
\begin{table}[!h]
    \centering
    \small
    \setlength{\tabcolsep}{6pt}
    \renewcommand{\arraystretch}{1.05}
    \begin{tabular}{lcccc}
        \toprule
        \textbf{Feature} & \textbf{Clean} & \textbf{Attack} & \textbf{Strong} & \textbf{Weak} \\
        \midrule
        BoW & 0.41& \textbf{0.69} & 0.87 & 0.23 \\
        \bottomrule
    \end{tabular}
    \caption{Average thinking-likeness scores.}
    \label{tab:style_likeness}
\end{table}

\subsection{Suffix Examples}
A few suffix examples optimized by $R^2A$ for each target router are showed  below.

{

    \noindent\textit{• RouteLLM--MF:}
        \begin{gadgetbox}
        hua jadx parece retorno Omar ※ ile Qty\_trans vào
        \end{gadgetbox}
   \noindent\textit{• RouteLLM-BERT:}
        \begin{gadgetbox}
        \}quanadermirroralsosedger\textbackslash ufffdofinemet\}translate\\9byteimpintegration`tonshan
        \end{gadgetbox}
    \vspace{2em}
    \noindent\textit{• RouteLLM-SW:}
        \begin{gadgetbox}
        shanãĥĶĠParsingĠsentenceenglĠtransformho\%times
        sort(graflangml(pmjanlangfilterĠGenerateĠin269arsi
        \end{gadgetbox}
    
    \noindent\textit{• RouteLLM-Causal:}
        \begin{gadgetbox}
            had !" !"ki\textbackslash ufffdwatprivacywill+no
        \end{gadgetbox}
    
    \noindent\textit{• P2L:}
        \begin{gadgetbox}
        originalrobot<*dad\{@estefuckoslivellanal\\naturedenhetcupshopimitray
        \end{gadgetbox}
    
    \noindent\textit{• Graph-Router:}
        \begin{gadgetbox}
        ]++eachbotatabamarmovieverbavieminhead\}\\\&solutionclick\textbackslash  ufffd`attend
        \end{gadgetbox}
    
    \noindent\textit{• RouterDC:}
        \begin{gadgetbox}
        summercanStyleType\&kokriadGamerGOtype\\File\{\{xorPhotWIAppaultureat
        \end{gadgetbox}
    \noindent\textit{• OpenRouter:}
        \begin{gadgetbox}
        decipherthispoem.alphabetical\}Yi-encodedoutput
        \end{gadgetbox}

}


\section{Ethical and Security considerations}
While our attack is intended to surface vulnerabilities in LLM routing and to inform the design of stronger defenses, it could be misused to inflate providers' inference costs or to bypass pricing tiers. 
We therefore recommend deploying routing-specific monitoring, rate limiting, and anomaly detection before exposing cost-aware routers in production, and we plan to release our triggers and code in a controlled manner to support defensive research rather than indiscriminate exploitation.

\end{document}